# A Novel Session Based Dual Steganographic Technique Using DWT and Spread Spectrum

## Tanmay Bhattacharya[*], Nilanjan Dey[**] and S. R. Bhadra Chaudhuri[***]

[*]Asst. Professor Dept. of IT, JIS College of Engineering, Kalyani, West Bengal, India.
[**]Asst. Professor Dept. of IT, JIS College of Engineering, Kalyani, West Bengal, India.
[***]Professor, Dept of E&TCE, Bengal Engineering and Science University Shibpur, Howrah, West Bengal, India.

## ABSTRACT

**This paper proposed a DWT based Steganographic technique. Cover image is decomposed into four sub bands using DWT. Two secret images are embedded within the HL and HH sub bands respectively. During embedding secret images are dispersed within each band using a pseudo random sequence and a Session key. Secret images are extracted using the session key and the size of the images. In this approach the stego image generated is of acceptable level of imperceptibility and distortion compared to the cover image and the overall security is high.**

## I. INTRODUCTION

*Steganography* [1, 2, 3] is the process of hiding of a secret message within an ordinary message and extracting it at its destination. Anyone else viewing the message will fail to know that it contains secret/encrypted data. The word comes from the Greek word "*steganos*" meaning "covered" and "*graphei*" meaning "writing".

LSB [4] insertion is a very simple and common approach to embedding information in an image in special domain. The limitation of this approach is vulnerable to every slight image manipulation. Converting image from one format to another format and back could destroy information secret in LSBs. Stego-images can be easily detected by statistical analysis like histogram analysis. This technique involves replacing N least significant bit of each pixel of a container image with the data of a secret message. Stego-image gets destroyed as N increases. In frequency domain data can be made secret by using Discrete Cosine Transformation (DCT) [5, 8]. Main limitation of this approach is blocking artefact. Grouping the pixel into 8x8 blocks and transforming the pixel blocks into 64 DCT co-efficient each. A modification of a single DCT co-efficient will affect all 64 image pixels in that block. One of the modern techniques of Steganography is Discrete Wavelet Transformation (DWT) approach [6, 7]. In this approach the imperceptibility and distortion of the Stego image is acceptable and it is resistant to several attacks.

## II. DISCRETE WAVELET TRANSFORMATION

The wavelet transform describes a multi-resolution decomposition process in terms of expansion of an image onto a set of wavelet basis function. The wavelet transform describes a multi-resolution decomposition process in terms of expansion of an image onto a set of wavelet basis functions. Discrete Wavelet Transformation has its own excellent space frequency localization properly. Applying DWT in 2D images corresponds to 2D filter image processing in each dimension. The input image is divided into 4 non-overlapping multi-resolution sub-bands by the filters, namely (LL1), (LH1), (HL1) and (HH1). The sub-band (LL1) is processed further to obtain the next coarser scale of wavelet coefficients, until some final scale "N" is reached. When "N" is reached, we'll have 3N+1 sub-bands consisting of the multi-resolution sub-bands (LLN) and (LHX), (HLX) and (HHX) where "X" ranges from 1 until "N". Generally most of the Image energy is stored in these sub-bands.

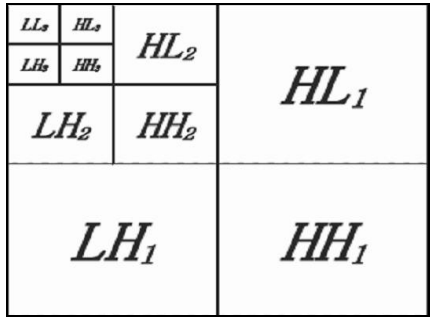

Fig.1 Three phase decomposition using DWT.

The Forward Discrete Wavelet Transform is very suitable to identify the areas in the cover image where a secret image can be embedded effectively due to its excellent space-frequency localization properties. In particular, this property allows the exploitation of the masking effect of the human visual system such that if a DWT co-efficient is modified, it modifies only the region corresponding to that coefficient. The embedding secret image in the lower frequency sub-bands ($LL_X$) may degrade the image significantly, as generally most of the Image energy is stored in these sub-bands. Embedding in the low-frequency sub-bands, however, could increase robustness significantly. In contrast, the edges and textures of the image and the human eye are not generally sensitive to changes in the high frequency sub-bands ($HH_X$). This allows the stego-image to be embedded without being perceived by the human eye. The compromise adopted by many DWT based algorithms, to achieve acceptable performance of imperceptibility and robustness, is to embed the secret image in the middle frequency sub-bands ($LH_X$) or ($HL_X$) and ($HH_X$). The Haar wavelet is also the simplest possible wavelet. Haar wavelet is not continuous, and therefore not differentiable. This property can, however, be an advantage for the analysis of signals with sudden transitions.

## III. CODE DIVISION MULTIPLE ACCESS (CDMA) SPREAD-SPECTRUM TECHNIQUE

Spread-spectrum technique can be described as a method in which a signal generated in a particular bandwidth when deliberately spread in the frequency domain, results in a signal with a wider bandwidth. If distortion is introduced in this signal by some process such as noise or filtering which damages only certain bands of frequencies, the message will be still in a recoverable state. In spread spectrum communications, the signal energy inserted into any one frequency is too undersized to create a visible artefact and the secret image is scattered over a wide range of frequencies, that it becomes robust against many common signal distortions. Because of its good correlation properties, noise like characteristics, easier to generate and resistance to interference, Pseudo noise sequences are used for Steganography.

## IV. PROPOSED ALGORITHM

*Secret Image Hiding:*

1. Cover image is decomposed into four sub bands (LL, LH, HL and HH) using DWT.

2. Two secret images are taken and converted into two different 1D Vectors.

3. Two different pseudo random 2D sequences are generated by the session based key.

4. Each HL and HH sub band of the cover image are modified separately using corresponding PN sequence depending upon the content of the corresponding secret 1D image vector to be embedded.

5. Four sub bands including two modified sub bands are combined to generate the stego image using IDWT.

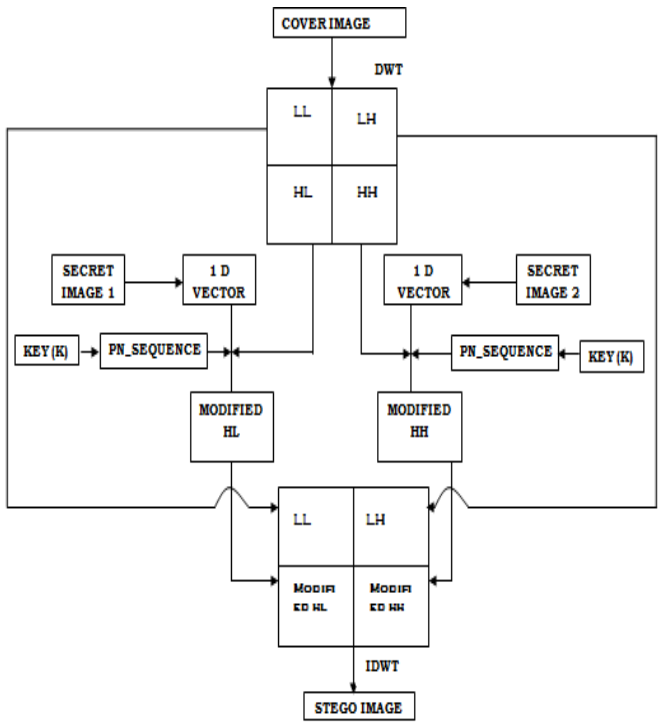

Fig 2: Image Hiding Process

*Secret Image Extraction*

1. Session key and Sizes of the secret images are sent to the intended receiver via a secret communication channel.

2. Secret images can recovered from the stego image using Correlation function and knowing the size of the secret image.

3. Extracted Secret Images are filtered to remove the unwanted signal.

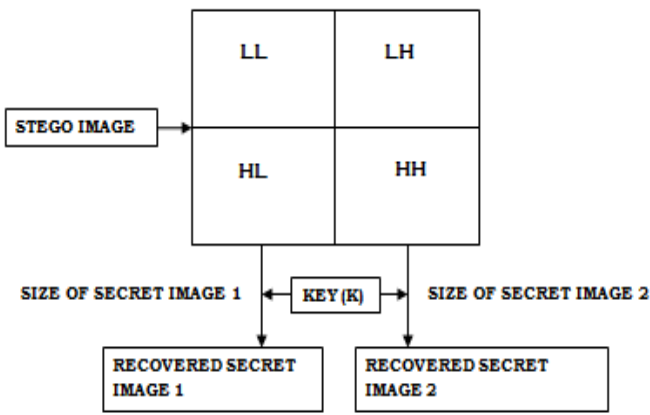

Fig 3: Image Extraction Process

## V. EXPLANATION OF THE ALGORITHM

*Secret Image Hiding*

Using DWT the Cover image is decomposed into four sub bands (LL, LH, HL and HH).

Two binary images, Secrete Image 1 and Secrete Image 2 are taken and converted into 2 one dimensional vectors.

Two pseudo random sequences are generated using a session based key and the size of any sub bands of the cover image.

Each of the bits of the binary secret image1 and binary secret image2 are embedded in HL and HH sub-bands respectively depending upon the elements of the one dimensional vector and the pseudo random sequences. The general equation used to embed the secret image is:

$$I_S(x, y) = I(x, y) + k \times S(x, y) \ldots\ldots\ldots (1)$$

In which $I(x, y)$ representing the selected DWT sub band of the cover image, $I_S(x, y)$ is the modified cover image, K denotes the amplification factor that is usually used to adjust the invisibility of the secrete images in corresponding sub bands. $S(x, y)$ is the pseudo random sequences.

Taking all the sub bands including the modified HH and LH sub bands, stego image is obtained applying IDWT (Inverse Discrete Wavelet Transformation)

*Secret Image Extraction procedure*

The session key and the size of the secrete images are provided to the intended receiver through a secrete communication channel.

The sub-bands are selected into which the secret image was embedded after applying DWT on stego image. The pseudo random sequence (PN) is generated using the same session based key which was used in the secret image embedding procedure. The correlation between the selected stego sub-band and the generated pseudo random sequence is calculated. Each correlation value with the mean correlation value is compared. If the calculated value is greater than twice the mean, then the extracted watermark bit will be taken as a 0, otherwise it is taken as a 1. The recovery process then iterates through the entire PN sequence until all the bits of the Secret image1 and secret image 2 have been recovered.

Filter is used on recovered secrete images to remove unwanted signals.

## VI. RESULTS

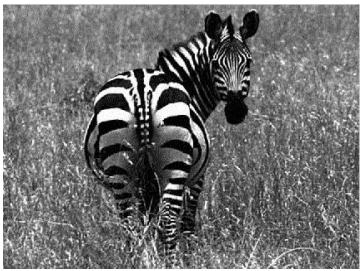

Fig. 4: Cover Image

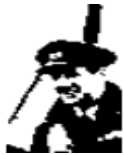

Fig. 5: Secret Image-1

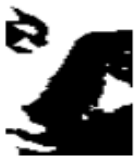

Fig. 6: Secret Image2

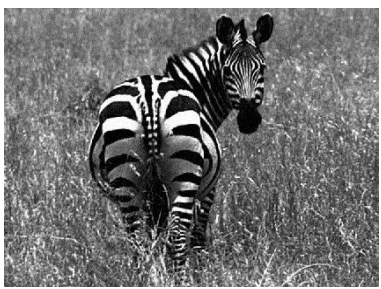

Fig. 7: Stego Image

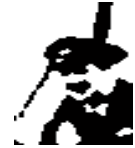

Fig. 8: Recovered Secret Image -1

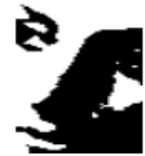

Fig. 9: Recovered Secret Image -2

Peak Signal to Noise Ratio (PSNR)

It measures the quality of a stego image. This is basically a performance metric and use to determine perceptual transparency of the stego image with respect to host image:

$$PSNR = \frac{MN \max_{x,y} P_{x,y}^2}{\sum_{x,y} (P_{x,y} - \overline{P}_{x,y})^2} \quad (1)$$

Where, M and N are number of rows and columns in the input image,

$P_{x,y}$ Is the original image and

$\overline{P}_{x,y}$ Is the Stego Image.

PSNR between Cover Image and Stego Image is 27.3850 shown in Table1.

| Cover Image vs. Stego Image | PSNR |
|---|---|
|  | 27.3850 |

Table 1

Correlation coefficient

After secret image embedding process, the similarity of original cover image x and stego images x' was measured by the standard correlation coefficient as follows:

$$Correlation = \frac{\sum (x-x')(y-y')}{\sqrt{(x-x')^2}\sqrt{(y-x')^2}} \quad (2)$$

Where y and y' are the discrete wavelet transforms of x and x'

Correlation between the secrete image1 and recovered secrete image1 after applying filter is 0.9381 and between the secrete image2 and recovered secrete image2 is 0.8870 shown inTable2.

| Correlation between original secret image and recovered secret image | Image1 | Image 2 |
|---|---|---|
| | 0.9381 | 0.8870 |

Table 2

## VII. CONCLUSION

In the proposed method second image is embedded in the HL sub band of the cover image. So there is a small visual change in between cover image and stego image. But due strong security aspects this small amount of imperceptibility is acceptable. This approach can be applied for colour image and audio Steganography also because DWT is applicable for any digital signal.